\documentclass[conference]{IEEEtran}
\IEEEoverridecommandlockouts
\usepackage{cite}
\usepackage{amsmath,amssymb,amsfonts}
\usepackage{algorithmic}
\usepackage{graphicx}
\usepackage{textcomp}
\usepackage{xcolor}
\usepackage{multirow} 
\usepackage{nicematrix} 
\def\BibTeX{{\rm B\kern-.05em{\sc i\kern-.025em b}\kern-.08em
    T\kern-.1667em\lower.7ex\hbox{E}\kern-.125emX}}
\begin{document}

\title{Energy-aware Incremental OTA Update for Flash-based Batteryless IoT Devices}

\author{
    \IEEEauthorblockN{Wei Wei\IEEEauthorrefmark{1}, Jishnu Banerjee\IEEEauthorrefmark{1}, Sahidul Islam\IEEEauthorrefmark{1}, Chen Pan\IEEEauthorrefmark{2}, Mimi Xie\IEEEauthorrefmark{1}}
    \IEEEauthorblockA{\IEEEauthorrefmark{1}Department of Computer Science, The University of Texas at San Antonio}
    \IEEEauthorblockA{\IEEEauthorrefmark{2}Department of Electrical \& Computer Engineering, The University of Texas at San Antonio}
    \small E-mail: \IEEEauthorrefmark{1}\IEEEauthorrefmark{2}\{wei.wei2, sahidul.islam, jishnu.banerjee, chen.pan, mimi.xie\}@utsa.edu \\
}

\maketitle

\begin{abstract}
Over-the-air (OTA) firmware updates are essential for updating and maintaining IoT devices, especially those batteryless devices reliant on energy harvesting power sources. Flash memory, favored for its low cost and high density, is extensively used for data storage in many IoT devices. However, due to its high energy demands for update operations, there is often insufficient energy for code updates. This paper proposes an incremental flash-based OTA update approach tailored for energy harvesting IoT devices, tackling the challenges brought by limited memory resources and fluctuating energy availability. Our approach is composed of three techniques: \emph{segment-based update packet design}, \emph{deferred flash segment writes}, and \emph{checkpoint-free update resumption}. \emph{Segment-based update packet design} segments firmware updates into smaller packets, each tailored for specific memory segments, thereby minimizing unnecessary flash operations and conserving energy. \emph{Deferred flash segment writes} accumulate packets in Static Random-Access Memory (SRAM) for collective processing, reducing the frequency of energy-intensive operations. Crucially, our \emph{checkpoint-free update resumption} ensures efficient recovery from power interruptions without significant energy cost on data backup. Through thorough experimental evaluation, we have observed that our approach significantly reduces the total energy consumed during OTA updates, and decreases the average total update time in energy harvesting environments.
\end{abstract}


\section{Introduction}

OTA firmware updates are essential for maintaining the security, functionality, and performance of batteryless IoT devices. They enable the remote modification of device firmware to patch vulnerabilities, add new features, or improve efficiency without requiring physical access to the devices. Considering the environmental and economic concerns associated with traditional battery-powered solutions, which often require onsite technical support or manual updates by manufacturers, the importance of OTA update capabilities grows when considering the maintenance and replacement costs in remote or extensive deployments. As batteryless IoT devices become more prevalent in a variety of settings, from urban infrastructures to isolated rural monitoring stations \cite{zhang2024intelligent}, the ability to update these devices remotely is crucial. In such scenarios, where manual updates are unfeasible or impossible, OTA updates ensure that batteryless IoT devices remain secure, functional, and up-to-date, regardless of their location.

Limited memory resources present a significant challenge for OTA updates in batteryless IoT devices. Traditional OTA update approaches typically require considerable memory overhead to store the entire new firmware image before installation, which often exceeds the memory capacity of the devices. To address this, researchers have developed more memory-efficient approaches such as incremental updates (IN) \cite{arakadakis2021firmware} or in-place patching \cite{zhang2016live}. These approaches are carefully designed to fit within tight memory limits but often introduce additional complexity and overhead during the update process. 

The fluctuating characteristics of the ambient energy sources, including solar, thermal, kinetic, or RF energy \cite{sun2023requirements}, introduce challenges in ensuring the stable operation of batteryless IoT devices, particularly for power-demanding processes such as OTA code updates. The energy demands associated with communication and flash operations may exceed the output of the energy harvesting power supply, resulting in incomplete updates that potentially compromise the device’s functionality and security.

To tackle the aforementioned challenges, this paper introduces a new energy-aware incremental OTA update (EA) approach for conducting OTA firmware updates specifically designed for flash-based batteryless IoT devices. It's important to emphasize that our approach concentrates on enhancing the performance of OTA updates at the software level, independent of hardware \cite{liu2023light} enhancements. Our approach includes the following contributions: 

\begin{itemize}
\item \emph{Segment-Based Update Packet Design}: Each update packet is designed to transmit update data exclusively for a designated flash memory segment. This precise, segment-focused design ensures that updates can be conducted on a packet-by-packet basis independently, optimizing the use of limited memory and energy. 

\item \emph{Deferred Flash Segment Writes}: In instances where update content for a single flash segment surpasses the maximal capacity of an update packet, the corresponding update packets are temporarily stored in the device's  SRAM pending the reception of all requisite packets for that segment. Upon completion, the accumulated packets are written to the flash memory through a flash erase and write operation. 

\item \emph{Checkpoint-Free Update Resumption}: This strategy bypasses the traditional need for energy-intensive flash-based checkpointing to save progress during the update process. Instead, the update distributor tracks the number of packets successfully received and processed by the IoT device. If a power failure interrupts the update, upon recovery, the device communicates with the update distributor to determine the last packet. 
\end{itemize}

\section{Background and Motivation}

\subsection{Background}

Flash memory, recognized for its cost-effectiveness and widespread adoption as a non-volatile storage solution, is essential to the architecture of batteryless IoT devices. It retains data using electrical charges in memory cells, preserving information without the need for continuous power. Each cell includes a control gate and a floating gate underneath, which is capable of being fully charged or discharged. The erasing and writing of the floating gate are performed through the oxide layer with high energy. The charge level of the floating gate alters the transistor's threshold, dictating the output of a logical 1 or 0 during the read operation.

\begin{figure}[htbp]
    \centering
    \includegraphics[width=88mm,height=29mm]{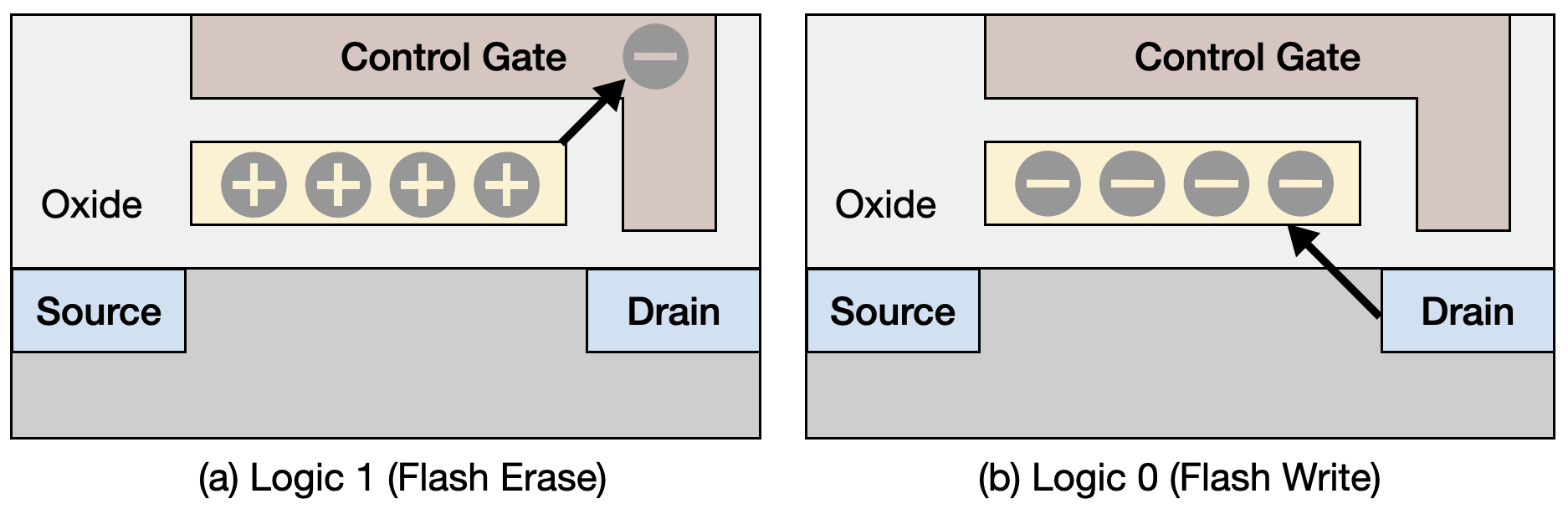}
    \caption{NOR Flash Memory Cell During Erase and Write Operations \cite{ti_flash}}
    \label{fig_flash_memory}
\end{figure}

The data writing process in flash memory is a two-step procedure, comprising erase and write. In this context, we utilize MSP430 NOR flash memory from our experiment as an illustrative example, as depicted in Fig. \ref{fig_flash_memory}. The erase process, illustrated in Fig. \ref{fig_flash_memory} (a), involves the application of a positive charge to the floating gate, effectively setting the cells within a segment to a default state of logic 1. Conversely, writing a flash cell negatively charges the floating gate, as illustrated in Fig. \ref{fig_flash_memory} (b). This step involves selectively setting certain bits within the segment to logic 0, correlating with the incoming data \cite{ti_flash}. The requirement for a segment-based erase before writing imposes a considerable energy demand, particularly significant during OTA firmware updates in batteryless IoT devices that operate within energy harvesting environments.

NOR flash memory emerges as the optimal choice for batteryless IoT devices due to its lower energy consumption and capability for direct code execution. It provides rapid read speeds along with comprehensive address and data buses, enabling both random and sequential data access, making it ideal for firmware storage in batteryless IoT devices that demand efficiency and direct access. 

\subsection{Motivation}

Traditional whole image update approaches often involve transmitting the entire firmware image to IoT devices, requiring storage with either flash memory or SRAM. One example of this approach is Live Firmware Update supported by Texas Instruments \cite{ti_flash_update}. This approach uses a duplicated flash memory bank to enable real-time OTA updates, reserving one bank exclusively for updates. Batteryless IoT devices, however, are typically designed with memory constraints to keep unit costs low for mass deployment. Doubling flash memory solely for update purposes increases unit costs without benefiting regular device operations, while relying on SRAM restricts update sizes to the SRAM's capacity and is not cost-effective due to SRAM's higher price compared to flash memory \cite{banerjee2021memory}. These considerations highlight the need for innovative OTA update approaches that accommodate the unique constraints of batteryless IoT devices, optimizing both resource utilization and cost-effectiveness.

Although incremental update approaches utilizing Ferroelectric Random Access Memory (FRAM) have been extensively researched \cite{zhang2016live}, \cite{wei2022intermittent}, \cite{arakadakis2021firmware}, their application to flash-based memory systems poses challenges, often triggering device shutdowns. This issue arises because each update packet may carry multiple delta differences, with each difference potentially affecting a distinct flash segment, leading to numerous flash write operations within a single packet update. Such frequent segment writes significantly heighten the risk of power outages. Moreover, in scenarios where energy harvesting is involved, adopting a packet-by-packet update strategy appears promising due to the potential for power outages. However, this strategy tends to distribute the update content for a single flash segment across various packets, substantially increasing flash write overhead. This scenario underscores the need for more sophisticated update mechanisms.

OTA communication operations along with flash memory operations, including erase and write, are the main factors driving energy consumption in flash-based batteryless IoT devices. These operations are notably energy-intensive and can rapidly deplete the available energy, potentially leading to power failures. Given their nature as non-interruptible processes that cannot easily save and restore their state, it is not feasible to pause and save the state of these operations halfway for later resumption. Consequently, any energy spent on these unsuccessful operations is effectively lost. When the device regains energy and turns back on, these operations must initiate from the beginning. This highlights the importance of developing an energy-efficient update approach that carefully considers the energy consumption of both communication and flash memory operations, ensuring updates can proceed with minimum energy waste.

\section{Energy-Aware Incremental OTA Update}

\subsection{Overview}

\begin{figure}[htbp]
    \centering
    \includegraphics[width=65mm,height=65mm]{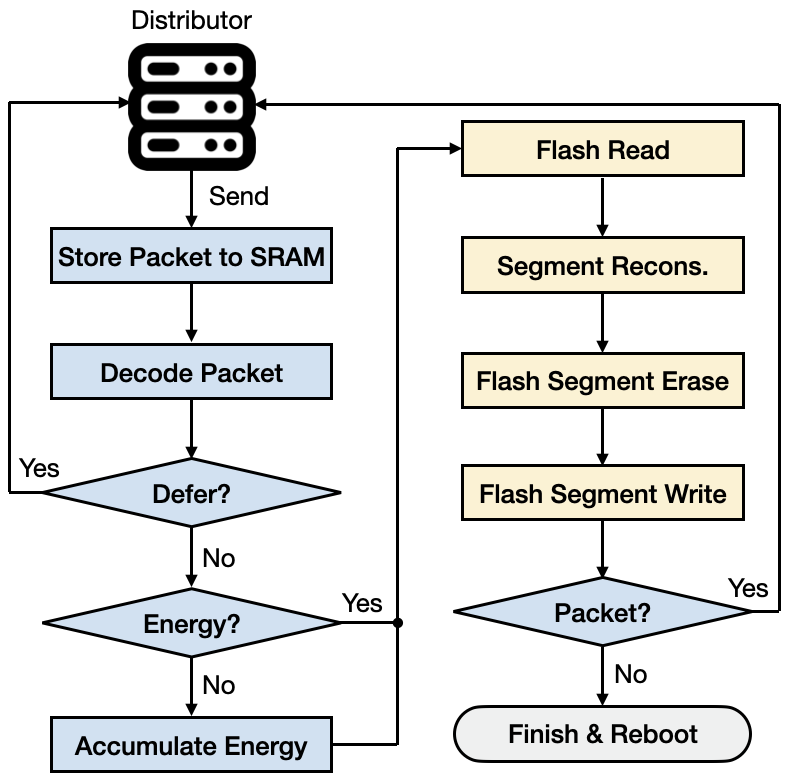}
    \caption{The Overview of the Proposed Approach}
    \label{fig_overview}
\end{figure}

In this paper, we introduce an energy-aware incremental OTA update approach for flash-based batteryless IoT devices, the details of which are illustrated in Fig. \ref{fig_overview}. The process initiates with the update distributor sending a notification to the IoT device, which, in turn, signals the distributor to send the update packet. Upon reception, the device stores the packet in SRAM and proceeds to decode it. 

Upon decoding the update packet, the device evaluates whether a defer flag has been set, which indicates if the writing to the flash memory should be delayed. In the absence of a defer flag, the device first reads the update-related original flash segment into SRAM. It then integrates the update delta from the decoded packet to reconstruct the updated segment. Following this reconstruction, the device erases the relevant flash segment and writes the reconstructed segment to the flash memory. This procedure is iteratively applied to each packet, intermittently stopping to gather the necessary energy.

If the defer flag is activated, the device postpones the flash operations until it has successfully stored all the update packets for the associated flash segment in SRAM. Once the device has received all necessary update packets, it proceeds with a sequence of operations: reading the existing flash segment, reconstructing the segment, erasing the flash segment, and then writing to the flash memory. This deferred approach minimizes the number of write operations needed for each flash segment, thus enhancing energy efficiency. The update process is completed once the device has processed all incoming packets, 

\subsection{Segment-Based Update Packet Design}

\begin{figure}[!htbp]
  \centering
  \includegraphics[width=88mm,height=22mm]{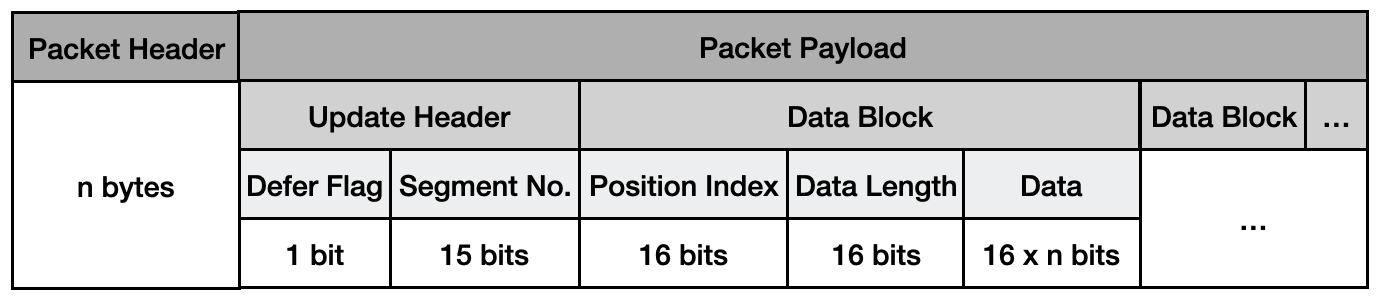}
  \caption{Segment-Based Update Packet Design Payload Structure}
  \label{fig_packet}
\end{figure}

To enhance the energy efficiency of incremental updates and make them suitable for batteryless IoT devices reliant on energy harvesting and flash memory, this paper introduces a \emph{segment-based update packet design}. This design, exemplified as the data structure of the packet payload depicted in Fig. \ref{fig_packet}, encapsulates update information for a single flash segment within each packet. Consequently, each update packet operates independently, eliminating the need for multiple flash operations during one packet update. 

Incorporated within the update header is a defer flag, designed to minimize write operations to the same flash segment. The specifics of the update packet, including the delta differences, are encapsulated within update blocks. These blocks can be decoded by the IoT device with the help of position indexes and data lengths scripts. Upon decoding an update packet, the device can reconstruct the updated content by combining the original content reading from the corresponding flash segment. This strategy significantly lowers the energy consumption per packet update, making it more suited to the fluctuating nature of energy harvesting power sources.

\subsection{Deferred Flash Segment Writes}

\begin{figure}[htbp]
    \centering
    \includegraphics[width=88mm,height=45mm]{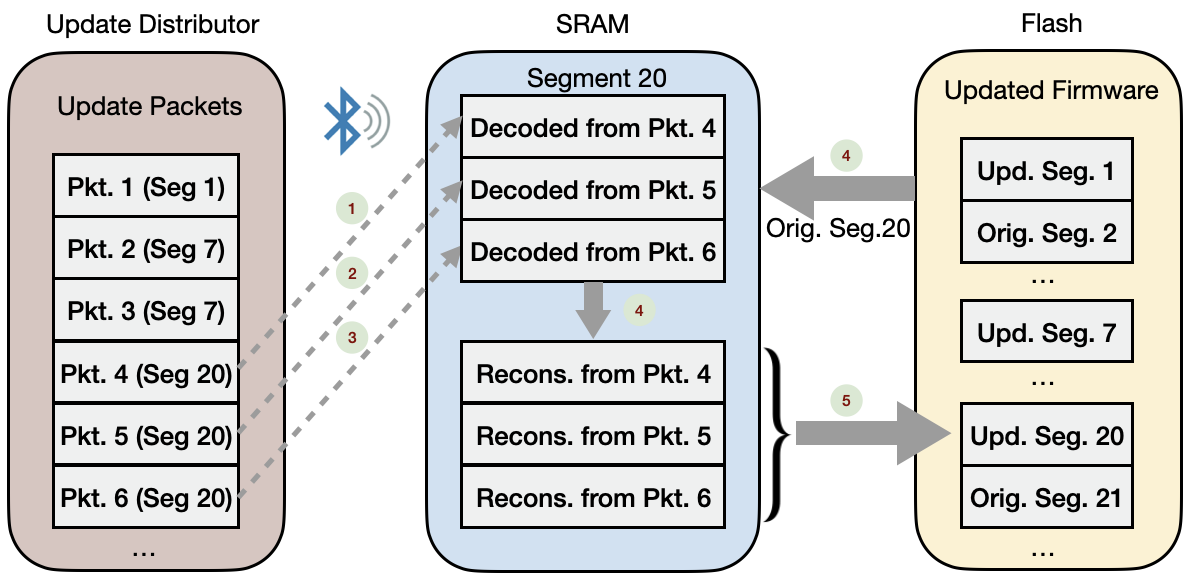}
    \caption{The Steps of Deferred Flash Segment Writes}
    \label{fig_defer}
\end{figure}

\emph{Deferred Flash Segment Writes} is a vital element of the energy-aware incremental OTA update approach, as detailed in Fig. \ref{fig_defer}. This strategy involves the following steps:

\begin{itemize}
\item Update packets are dispatched by the Update Distributor, with each packet corresponding to a specific flash memory segment. A single segment may require multiple packets for a full update; for instance, as depicted in Fig. \ref{fig_defer}, segment 1 needs one packet, while segment 20 requires three.

\item Upon receiving an update packet, the IoT device temporarily stores it in SRAM and decodes it. If a defer flag is detected in the packet header, the device signals the update distributor for the additional packets associated with the same flash segment. This is exemplified in Fig. \ref{fig_defer} where packets for segment 20 are accumulated in steps 1 to 3. This temporary deferral is vital, as it prevents multiple flash operations to the same flash segment.

\item After receiving all associated packets for a segment, the IoT device reconstructs the complete updated segment in SRAM, using the decoded packets and the original data read from flash memory, as indicated in step 4 of Fig. \ref{fig_defer}.

\item Finally, in step 5, the fully reconstructed segment, segment 20 in this example, is written to the flash memory in a single write operation that includes an erase followed by a write.

\end{itemize}

\subsection{Checkpoint-Free Update Resumption}

\begin{figure}[!htbp]
  \centering
  \includegraphics[width=70mm,height=63mm]{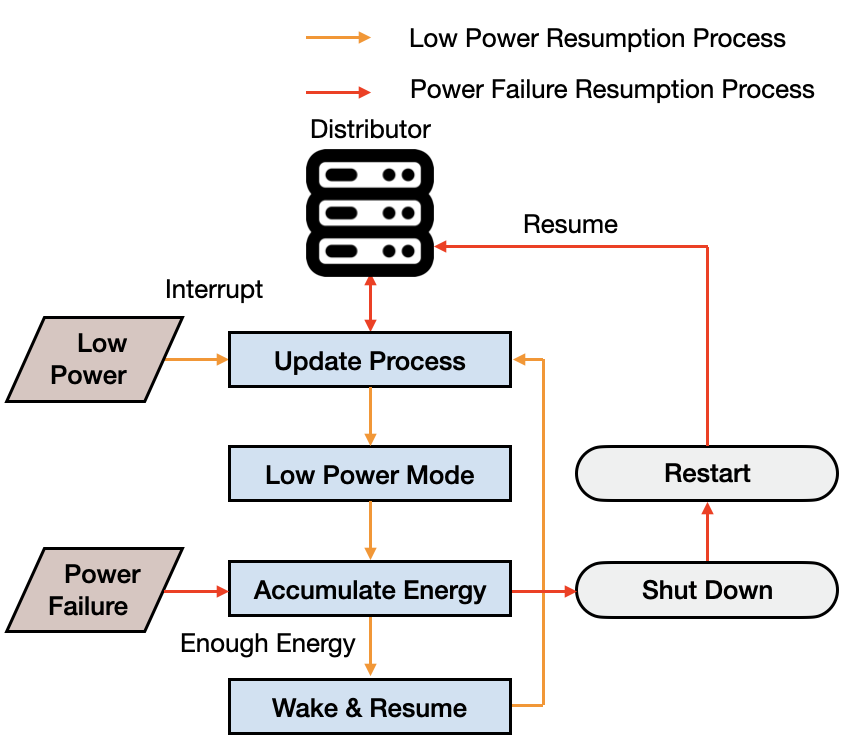}
  \caption{Update Resumption with  Power Constraints}
  \label{fig_resume}
\end{figure}

While entering low power mode can significantly lower the risk of power failures in IoT devices, it does not eliminate the possibility. Even in low power mode, maintaining data in SRAM consumes some energy. If the energy required for data retention in SRAM surpasses the amount being harvested, a power failure could still occur. Consequently, a mechanism to efficiently resume the update process after a power failure remains necessary.

The idea of using a checkpoint technique to safeguard update packets might seem practical at first glance. However, the energy cost for such backups, reaching up to 216 µJ, greatly surpasses the mere 65.6 µJ needed for retransmitting an update packet. This significant energy difference prompted us to devise a checkpoint-free strategy that relies on the update distributor's ability to track the update progress. Given that our approach updates firmware on a per-segment basis, the update distributor can accurately determine the number of segments successfully updated by the IoT device based on its requests. Thus, if an IoT device recovers from a power failure, the update distributor can resume the update process from the last unsuccessful segment. While this strategy may require the retransmission of some update packets (1 to 3 packets), it offers a viable compromise, enabling a reasonable level of resumption without the heavy energy costs of traditional checkpoint approaches. The update resumption process is depicted in Fig. \ref{fig_resume}, illustrating our approach's adaptability to various power constraints. 
\section{Experiments and Evaluation}

\subsection{Experimental Setup}
\begin{figure}[!htbp]
  \centering
  \includegraphics[width=75mm,height=80mm]{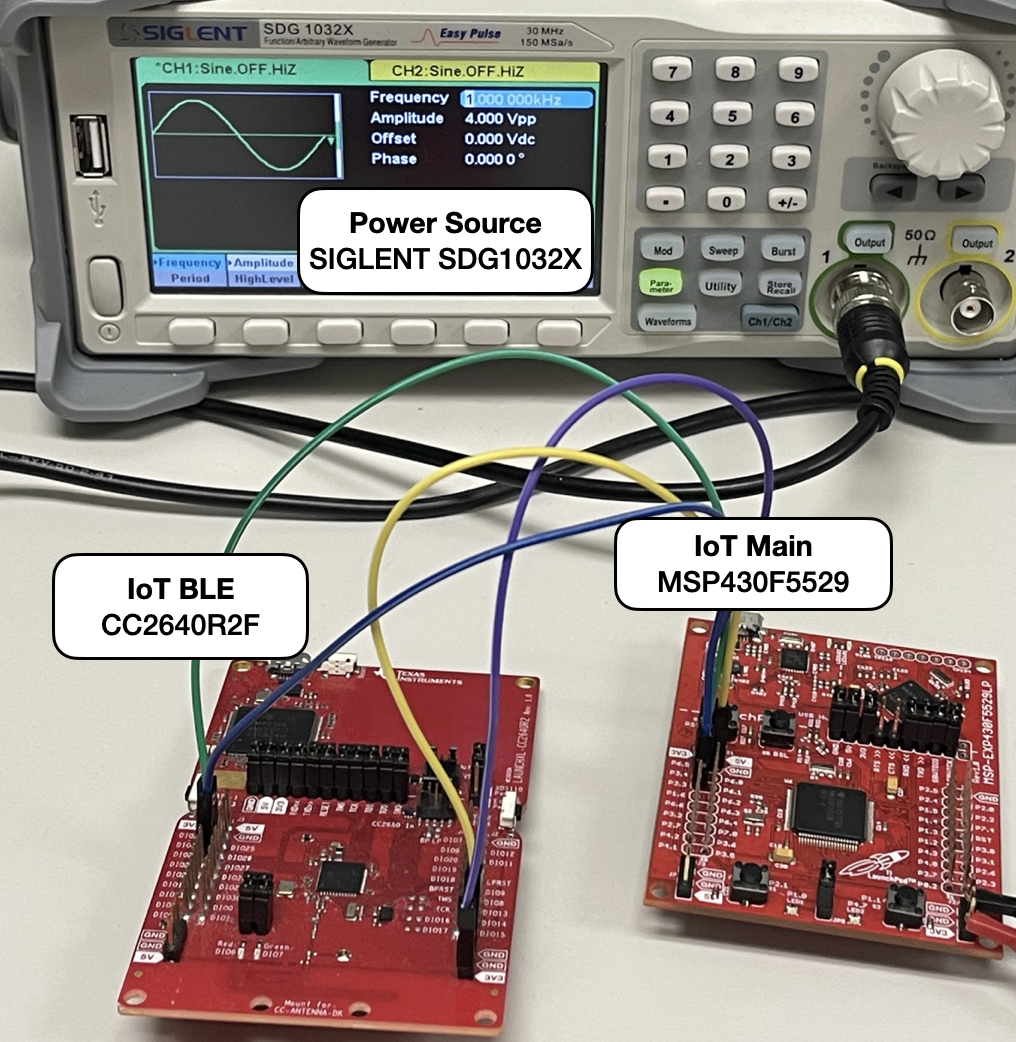}
  \caption{Hardware Setup}
  \label{fig_exp}
\end{figure}

Our experimental setup features an IoT device that integrates an MSP430F5529 \cite{msp430f5529} microcontroller with a CC2640R2F \cite{CC2640R2F} communication module. For data transmission within the IoT device, UART communication is used. The MSP430F5529 is a traditional flash-based IoT device, equipped with 128 KB of NOR flash memory and 8 KB of SRAM, and is configured to operate at 1 MHz to minimize energy consumption. The CC2640R2F module facilitates communication, employing the Bluetooth LE 5.0 protocol \cite{ble5} for OTA data transfer. To test various power conditions, we employ a Siglent SDG1032X function generator. The detailed hardware setup is shown in Fig. \ref{fig_exp}.

\subsection{Benchmark}

The selection of benchmarks for this evaluation was driven by the aim to include a broad spectrum of update tasks for batteryless IoT devices, such as enhancing security, introducing new functionalities, boosting performance, and resolving bugs. Each benchmark was carefully selected to closely simulate the operational challenges encountered by batteryless IoT devices in real-world scenarios, thereby offering a solid foundation for assessing the effectiveness of our approach. The benchmarks utilized in this study are outlined in Table \ref{table_benchmark}, which also includes the original sizes of the benchmarks for reference.

\begin{table}[htbp]
\centering
\caption{Benchmarks}
\label{table_benchmark}
\begin{tabular}{|| l | l | l ||}
\hline
Benchmark                    & Update Description               & Size (Bytes) \\ \hline\hline
MTH\cite{guthaus2001mibench} & Basic Math: add more operations  & 30814       \\ \hline
STR\cite{guthaus2001mibench} & String Search: small str to large str    & 8254        \\ \hline
SRT\cite{guthaus2001mibench} & Quick sort: add more sort items  & 57324       \\ \hline
AES\cite{taha2014key}        & AES Encryption: AES 128 to 256   & 3355        \\ \hline
OKG\cite{islam2022enabling}  & OKG model: weights update        & 56210       \\ \hline
\end{tabular}
\end{table}

\subsection{Evaluation}

In our analysis of the energy consumption for different OTA update stages, we utilize detailed measurements from \cite{liu2023light}, \cite{wei2022intermittent}. These state-of-the-art studies offer an exhaustive analysis of energy use throughout the OTA update process, providing essential insights that aid in evaluating our approach's efficiency. To have a more accurate evaluation, we incorporate the size of the packet header into the transmitted raw data (packet payload) and calculate the energy per byte to evaluate our approach's energy efficiency. The details of energy consumption are shown in table \ref{table_energy_consumption}. 

\begin{table}[htbp]
\centering
\caption{Energy Consumption Analysis of OTA Updates \cite{liu2023light}, \cite{wei2022intermittent}}
\label{table_energy_consumption}
\begin{tabular}{| l | l | l |}
\hline
Name        & Description                                 & Value       \\ \hline
$E_{b}$     & Everage energy cost per byte transmission   & 0.251 µJ    \\ \hline
$E_{e}$     & Erase one segment from flash                & 137.2 µJ    \\ \hline
$E_{w}$     & Write (after erase) one segment to flash    & 78.80 µJ    \\ \hline
$E_{r}$     & Read and reconstruct one segment from flash & 0.500 µJ    \\ \hline
$T_{b}$     & Everage time per byte transmission          & 9.361 µs    \\ \hline
$T_{e}$     & Time to erase one flash segment             & 27.00 ms    \\ \hline
$T_{w}$     & Time to write one flash segment (erased)    & 16.00 ms    \\ \hline
$P_{l}$     & Power consumption in low power mode         & 89.00 µW    \\ \hline
\end{tabular}
\end{table}

\subsubsection{Total Update Size}

We compare our Energy-Aware OTA Update (EA) approach with two other state-of-the-art approaches: Light Flash Write (LW) \cite{liu2023light} and Incremental Update (IN) \cite{wei2022intermittent}. For practical considerations, the LW approach is restricted to the AES benchmark due to the 8 KB SRAM limit. However, for the sake of this comparison, we hypothetically extend the SRAM capacity to accommodate the LW approach across all benchmarks. The specifics of this comparison are illustrated in Fig. \ref{table_update_size}.

The LW approach involves transmitting the entire updated firmware image to the IoT device's SRAM, followed by writing the complete image to the flash memory all at once. While this approach might result in a relatively low number of flash writes, as demonstrated in the AES update benchmark, it requires the transmission of the entire firmware, including unchanged parts. This could result in larger total data transmissions and additional flash operations compared to other approaches, particularly noticeable in scenarios like the SRT benchmark. A significant drawback of the LW is the requirement for temporary storage of all update packets in SRAM, constraining the update size to the SRAM's capacity, which is 8 KB in our experiments. Consequently, the LW approach fails to update four out of five benchmarks due to this restriction.

The IN approach sends update packets sequentially and updates the flash memory packet-by-packet. It successfully updates all five benchmarks with the smallest total update size. However, not being optimized for flash memory, results in the highest number of flash writes. When comparing the energy consumption of a single flash write operation (erase + write) at 216 µJ to the energy required to transmit one packet (261 bytes with header) at 65.6 µJ, the excess energy used for the numerous flash writes far exceeds the savings achieved from reduced data transmission. This discrepancy becomes especially evident in the MTH and STR benchmarks, highlighting the need for a more balanced approach.

The EA approach, our proposed approach, strategically balances the transmission size and the number of flash writes by employing a \emph{segment-based update packet design} and \emph{deferred flash segment writes}. Across all five benchmarks, the EA approach achieves a comparable low count of flash writes similar to the best of the other approaches, while slightly increasing the total update size in comparison to the IN approach.

The EA approach, our proposed approach, strategically balances the transmission size and the number of flash writes by employing a \emph{segment-based update packet design} and \emph{deferred flash segment writes}. Across all five benchmarks, the EA approach achieves a comparable low count of flash writes similar to the best of the other approaches, while slightly increasing the total update size in comparison to the IN approach.

\begin{table}[htbp]
\centering
\caption{Comparative Analysis of Update Sizes in Bytes}
\label{table_update_size}
\begin{tabular}{*{10}{c}}
\hline
& \multicolumn{3}{c}{Total Update Size} & \multicolumn{3}{c}{No. of Packets} & \multicolumn{3}{c}{No. of Writes}      \\ \hline  
BM  & LW    & IN    & EA    & LW  & IN  & EA  & LW  & IN  & EA  \\ \hline
MTH & 33890 & 28644 & 29627 & 130 & 111 & 146 & 62  & 171 & 62  \\ \hline
STR & 33085 & 34014 & 34337 & 127 & 131 & 134 & 61  & 189 & 61  \\ \hline
SRT & 61319 & 2497  & 3455  & 235 & 10  & 60  & 229 & 68  & 60  \\ \hline
AES & 3682  & 3059  & 3165  & 15  & 13  & 17  & 7   & 19  & 8   \\ \hline
OKG & 60137 & 10767 & 11108 & 231 & 42  & 53  & 110 & 60  & 19  \\ \hline
\end{tabular}
\end{table}

\subsubsection{Total Update Energy Consumption}

\begin{figure}[!htbp]
  \centering
  \includegraphics[width=88mm,height=60mm]{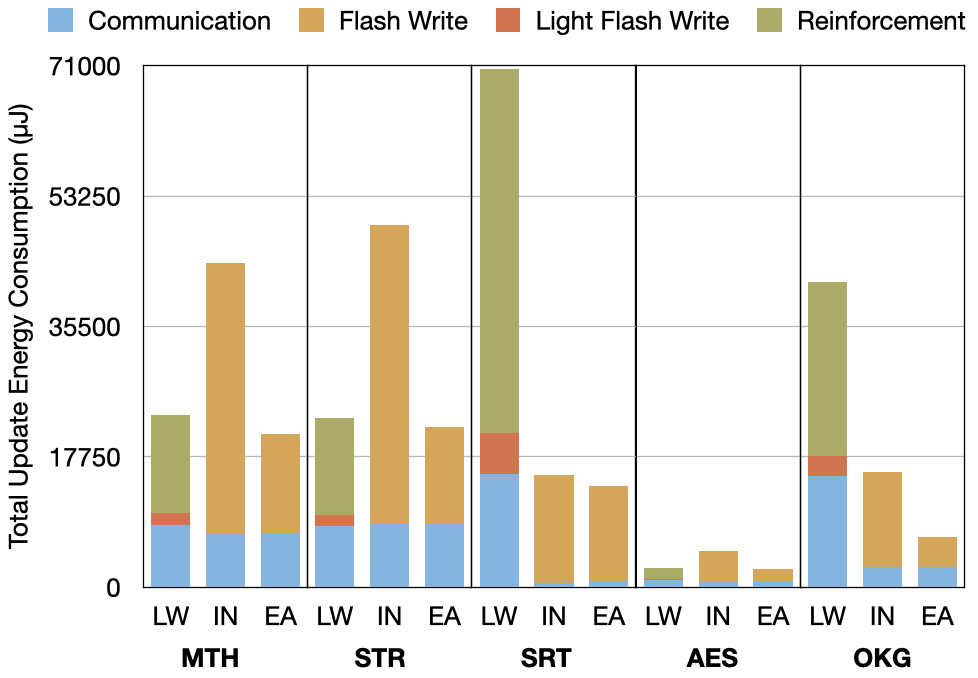}
  \caption{Total Update Energy Consumption}
  \label{fig_update_energy}
\end{figure}

To assess the energy efficiency of different OTA update approaches, we conducted measurements across our designed benchmarks. In Fig. \ref{fig_update_energy}, the energy consumption is divided into four components: communication, flash write, light flash write, and reinforcement. Light flash write and reinforcement are special techniques used in the LW. Although light flash write can dramatically reduce the energy for the updates, it still requires an equal amount of energy compared with the other two approaches for flash segment reinforcement. 

The LW approach, although limited in its applicability, achieves the highest energy efficiency in the AES benchmark context. Nevertheless, this efficiency diminishes with smaller update sizes, limiting its overall effectiveness.

The IN approach excels in communication efficiency but is hindered by the significant energy consumption of flash write, surpassing the energy savings from communication. This discrepancy leads to the IN approach having the highest overall energy consumption.

The EA approach effectively balances the costs of communication energy and flash operation, leading to the lowest energy consumption in 4 out of 5 benchmarks. Moreover, it matches the performance of the LW approach in scenarios where the LW is most efficient, demonstrating its adaptability and efficiency across a range of update situations.

\subsubsection{Everage Total Update Time}

\begin{figure}[!htbp]
  \centering
  \includegraphics[width=88mm,height=60mm]{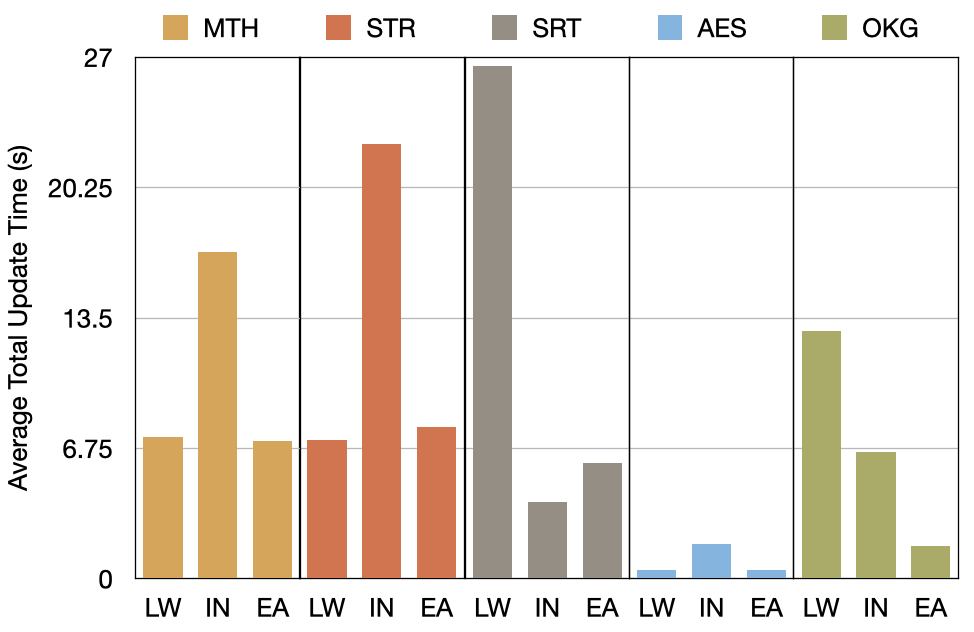}
  \caption{Average Total Update Time}
  \label{fig_update_time}
\end{figure}

To closely mimic real-world energy harvesting scenarios and evaluate the performance of our approaches in such environments, we developed a custom power simulator. This simulator employs a power trace derived from measurements of energy harvested from a real radio frequency WiFi device, showcasing a somewhat random but consistently patterned energy harvesting behavior. For a comprehensive analysis, we generated 1000 power traces, each representing a distinct energy harvesting scenario.

Our evaluation was conducted within this simulated energy harvesting environment, using a typical 400 mF capacitor for energy storage. To ensure an equitable comparison, the reinforcement time for the LW approach was excluded from the total update time calculation. This evaluation centered on determining the average total time required to complete the update process for each approach. The results of this analysis are detailed in Fig. \ref{fig_update_time}.

Although the LW approach significantly reduces the time required for flash write operations compared to the other two approaches, it does not outperform our EA approach in terms of average total update time, as illustrated in the STR and AES benchmarks. Specifically, in the SRT and OKG benchmarks, the LW approach demonstrates notably lower performance. Similarly, while the IN approach performs slightly better in the SRT benchmark due to its lower communication costs, it has considerably higher overhead in the remaining benchmarks. This analysis reveals the EA approach's superior efficiency across a broader spectrum of benchmarks, underlining its suitability in different energy harvesting scenarios.
\section{Conclusion}

In conclusion, this paper presents an innovative approach for OTA firmware updates tailored to the unique needs of flash-based batteryless IoT devices powered by energy harvesting. By employing a \emph{segment-based update packet design}, \emph{deferred flash segment writes}, and \emph{checkpoint-free update resumption}, we have significantly enhanced OTA update efficiency. Our comprehensive experimental analysis demonstrates that our approach not only substantially reduces energy consumption during OTA updates but also decreases the average time required to complete these updates. These results highlight the effectiveness of our approach and its potential to facilitate more sustainable and efficient OTA updates in energy harvesting environments.

\bibliographystyle{unsrt}
\bibliography{Bib/ref}

\end{document}